\documentclass[aps,pra,twocolumn,amsmath,amssymb,nofootinbib,superscriptaddress,floatfix,epsfig]{revtex4-1}
\usepackage{graphicx}
\usepackage{tabularx}
\usepackage{subfigure}
\usepackage{epsfig}
\usepackage{dcolumn}
\usepackage{bm}
\usepackage{euscript}
\usepackage{color}
\usepackage{epsfig,psfrag,subfigure}
\usepackage{amsfonts}
\usepackage{exscale}
\usepackage{amsbsy}

\def\avg#1{\left\langle#1\right\rangle}

\def\braket#1#2{\left\langle #1\right|\left.#2\right\rangle}

\def\be{\begin{equation}}       \def\ee{\end{equation}}
\def\bea{\begin{eqnarray}}      \def\eea{\end{eqnarray}}
\def\ba{\begin{array}}
\def\ea{\end{array}}
\def\bnum{\begin{enumerate} }
\def\enum{\end{enumerate}}

\def\nn{\nonumber}

\def\=>{\Rightarrow}
\def\>{\rightarrow}

\def\eye2{Fathbb{I}}

\def\Eq#1{Eq.~(\ref{#1})}
\def\Fig#1{Fig.~\ref{#1}}

\renewcommand{\>}{\rangle}

\usepackage{graphicx}
\usepackage{dcolumn}
\usepackage{bm}

\usepackage{braket}
\usepackage{mathrsfs}
\usepackage{hyperref}

\newcommand{\norm}[1]{\vert #1 \vert}
\renewcommand{\vec}[1]{\boldsymbol{#1}}

\begin{document}
\title{Fermion-induced quantum critical point in Dirac semimetals: a sign-problem-free quantum Monte Carlo study}
\author{Bo-Hai Li}
\affiliation{Institute for Advanced Study, Tsinghua University, Beijing 100084, China}
\author{Zi-Xiang Li}
\affiliation{Department of Physics, University of California, Berkeley, CA 94720, USA}
\affiliation{Materials Sciences Division, Lawrence Berkeley National Laboratory, Berkeley, CA 94720, USA}
\author{Hong Yao}
\affiliation{Institute for Advanced Study, Tsinghua University, Beijing 100084, China}%
\affiliation{State Key Laboratory of Low Dimensional Quantum Physics, Tsinghua University, Beijing 100084, China}%
\affiliation{Department of Physics, Stanford University, Stanford, California 94305, USA}
\date{\today}
\begin{abstract}
According to Landau criterion, a phase transition should be first order when cubic terms of order parameters are allowed in its effective Ginzburg-Landau free energy. Recently, it was shown by renormalization group (RG) analysis that continuous transition can happen at putatively first-order $Z_3$ transitions in 2D Dirac semimetals and such non-Landau phase transitions were dubbed ``fermion-induced quantum critical points''(FIQCP) [Li {\it et al.}, Nature Communications {\bf 8}, 314 (2017)].
The RG analysis, controlled by the 1/$N$ expansion with $N$ the number of flavors of four-component Dirac fermions, shows that FIQCP occurs for $N\geq N_c$. Previous QMC simulations of a microscopic model of SU($N$) fermions on the honeycomb lattice showed that FIQCP occurs at the transition between Dirac semimetals and Kekule-VBS for $N\geq 2$. However, precise value of the lower bound $N_c$ has not been established. Especially, the case of $N=1$ has not been explored by studying microscopic models so far. Here, by introducing a generalized SU($N$) fermion model with $N=1$ (namely spinless fermions on the honeycomb lattice), we perform large-scale sign-problem-free Majorana quantum Monte Carlo simulations and find convincing evidence of FIQCP for $N=1$. Consequently, our results suggest that FIQCP can occur in 2D Dirac semimetals for all positive integers $N\geq 1$.
\end{abstract}
\maketitle

\textbf{Introduction}: Universal behaviors of interacting many-body systems near quantum phase transitions are among central interest in modern condensed matter physics ~\cite{WenXGbook,Fradkinbook,Subirbook,Sondhi1997}. In  Landau-Ginzburg theory ~\cite{Landaubook}, a prevalent understanding of phase transitions is provided by order parameters whose non-zero expectation value characterizes spontaneously symmetry-breaking phases. Sufficiently close to the transition point, fluctuations of order parameters at low energy dominate and are described by a continuum field theory of order parameters. In combination with Wilson's renormalization group (RG) theory \cite{Wilson1974}, the Landau-Ginzburg-Wilson (LGW) paradigm has made seminal contributions to understanding continuous phase transitions in correlated many-body systems. Quantum criticality beyond the LGW paradigm, often called Landau-forbidden or non-Landau transitions, has attracted increasing attentions in the past few decades. 

Landau proposed two main criterions about under what conditions first-order transitions must occur. One criterion states that a transition between two phases with non-compatible broken symmetries should be first order. Nonetheless, continuous transitions violating this Landau criterion may be realized in certain many-body systems \cite{Senthil2004,Senthil2004prb,Levin2004,Motrunich2004,Levin2004,Akihiro2005,ChongWang2017,CenkeXu2018, DHLee2010,YizhuangYou2018} where fractionalized excitations at the transition are essential in rendering the transition from first order to second order. Such continuous non-Landau transitions were called ``deconfined quantum critical points'' (DQCP) \cite{Senthil2004}. Tremendous progress has been witnessed in looking for DQCP in models \cite{Assaad2016PRX,Qin2017PRX,Zhang-XF2018PRL, Sandvik2007,Kaul2008,Kaul2011,Sandvik2009,Troyer2008,Sandvik2012,TaoXiang2012,Prokof2013,Bartosch2013,FaWang2015,
Xiaogangwen2016,ZiyangMeng2018,WenanGuo2016,Assaad2017,Zaletel2018,WenanGuo2018,Nahum2015,Nahum20152}. Recently, large-scale QMC simulations \cite{Li2019DQCP} of microscopic fermion models obtained critical exponents that are consistent with the conformal bootstrap bounds \cite{Ginsparg1988,Poland2019Bootstrap,Bobev2015PRL}, providing further support of DQCP with emergent SO(5) symmetry.

The other Landau criterion states that continuous phase transitions are forbidden when cubic terms of order parameters are allowed in its effective Ginzburg-Landau (GL) free energy. For instance, the quantum three-state Potts model in 2+1D has been convincingly shown to possess a first-order quantum phase transition \cite{Wu1982} since the allowed cubic terms of its $Z_3$-order parameter in the low-energy GL free energy are relevant in the sense of RG. Recently, an intriguing scenario beyond this Landau cubic criterion was introduced [\onlinecite{FIQCP1}]: a putatively first-order transition in the sense of the Landau cubic criterion can be rendered into a continuous transition by coupling gapless Dirac fermions to fluctuations of $Z_3$ order parameters in its low-energy 2+1D GL theory. Such continuous transitions were dubbed as ``fermion-induced quantum critical points'' (FIQCP) \cite{FIQCP1,FIQCP2,FIQCP3,Herbut2018-FIQCP,Herbut2017-FIQCP,Herbut2016-FIQCP}. The RG analysis in \cite{FIQCP1}, controlled in the 1/$N$ expansion with $N$ being the number of flavors of four-component Dirac fermions in 2+1D, showed that FIQCP can occur when $N\geq N_c$, where $N_c$ being the lower bound of $N$ for FIQCP. Namely, in RG analysis, for $N$ larger than a critical value $N_c$, cubic terms of order-parameter field become irrelevant which results in a continuous quantum phase transition violating the Landau cubic criterion.

So far the precise value of $N_c$ remains unknown, although the previous large-$N$ RG analysis predicted that $N_c=0.5$ \cite{FIQCP1} and the FRG analysis obtained $N_c\approx 1.9$ \cite{Herbut2017-FIQCP}. It is challenging to obtain the precise value of $N_c$ from RG analysis. However, for integer $N$, it is possible to study such transitions in microscopic models by numerical methods such as quantum Monte Carlo simulations (QMC) \cite{Blankenbecler1981,Blankenbecler1981,Fucito1981,AssaadBook}. Ref. \onlinecite{FIQCP1} introduced a microscopic model of SU$(N)$ fermions on the honeycomb lattice and found a quantum phase transition between 2+1D gapless Dirac semimetal (DSM) \cite{Castro2009,Herbut2006Graphene,Novoselov2005,Roy2013PRB} to Kekule valence bond solids (Kekule-VBS) for $N\geq 2$ \cite{CongjunWu2016,CongjunWu2017,PALee2018, XiaoyanXu2019,YuanyaoHe2019,Pujari2013PRL,Ryu2009PRB,Hou2007PRL,Moessner2001}. The Kekule-VBS transition can be characterized by a $Z_3$ order parameter whose cubic terms are allowed in the GL free energy, implying a first-order transition according to the Landau cubic criterion. However, from large-scale QMC studies, it was shown convincingly that FIQCP occurs for $N\geq 2$ (specifically $N=2,3,4,5,6$). However, whether FIQCP may occur for the case of $N=1$ (the lowest possible integer value of $N$) remains open.

Note that the SU($N$) model introduced in Ref. \onlinecite{FIQCP1} does not support Kekule-VBS phase transition for $N=1$. To study the nature of $Z_3$ quantum phase transition in DSM with $N=1$, it is desired to construct a microscopic model which features a transition between DSM and Kekule-VBS for $N=1$. Here, we propose a generalized SU($N$) fermion model for which the transition between the DSM and Kekule-VBS can be realized down to $N=1$. Moreover, for this $N=1$ model of spinless fermions, QMC simulations can be made sign-problem-free using Majorana representation [\onlinecite{MQMC1,MQMC2,TaoXiang2016}]. As an intrinsically-unbiased approach, QMC is often employed to explore interacting quantum models that are sign-problem-free \cite{Troyer2005,CongjunWu2005,Berg2012,MQMC1,MQMC2,TaoXiang2016,LeiWang2015} (For a recent review, see, e.g. Ref. \onlinecite{LYQMCReview}). Consequently, we employ sign-problem-free Majorana quantum Monte Carlo (MQMC) simulations to investigate whether the Kekule-VBS transition features a FIQCP or not for the case of $N=1$. From our large-scale MQMC simulations, we find convincing evidences of a continuous quantum phase transition between the $N=1$ DSM and Kekule-VBS phases. At the critical point, the U(1) symmetry emerges indicating that the transition falls into the chiral XY universality class~\cite{Scherer2017,Sorella2018,Sorella2016,Rosenstein1993,Moshe2003}. Indeed, the critical exponents obtained from MQMC simulations are reasonably consistent with the ones obtained from previous RG analysis of chiral XY universality class. Consequently, we believe that FIQCP can occur at the transition between DSM and Kekule-VBS phase for the case of $N=1$.
As it was rigorously proved from supersymmetry that the transition cannot be continuous for $N=1/2$ \cite{FIQCP1,FIQCP3}, we conclude that the lower bound $N_c$ of four-component fermion flavors for realizing FIQCP should satisfy $1/2<N_c\leq1$.

\begin{figure}[t]
    \includegraphics[width=0.4\linewidth]{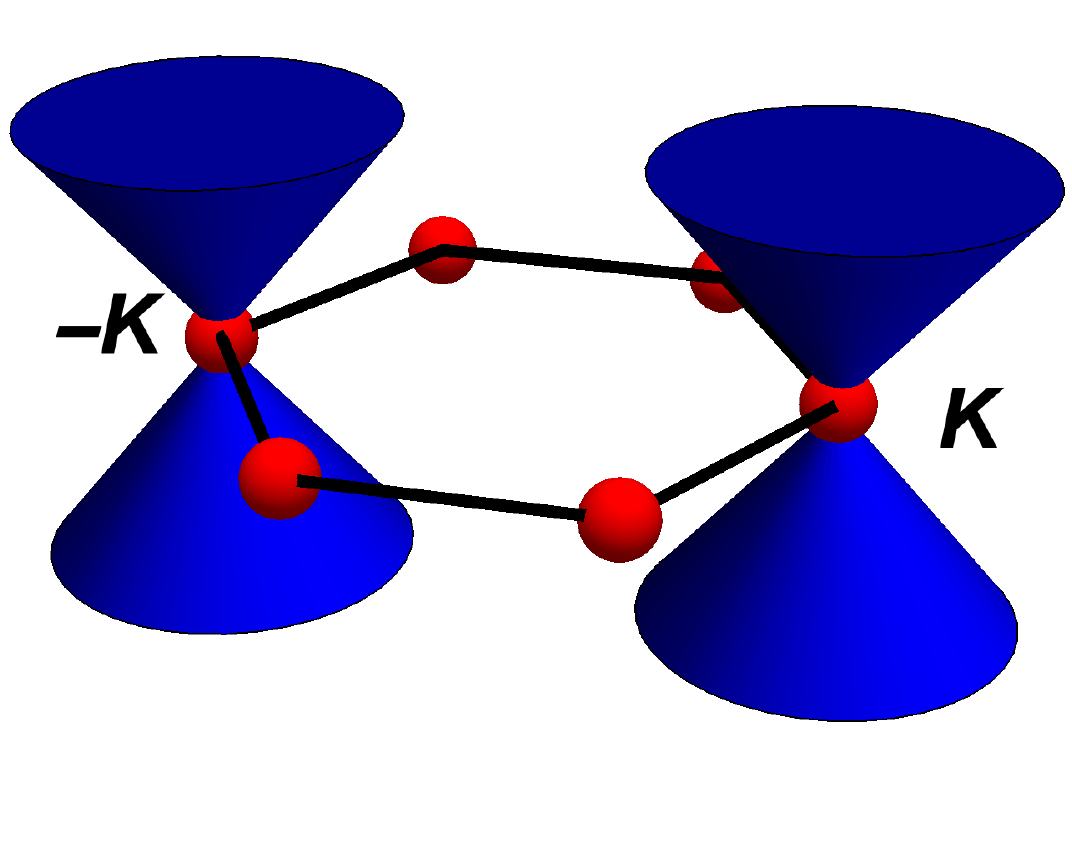}~~~~
    \includegraphics[width=0.4\linewidth]{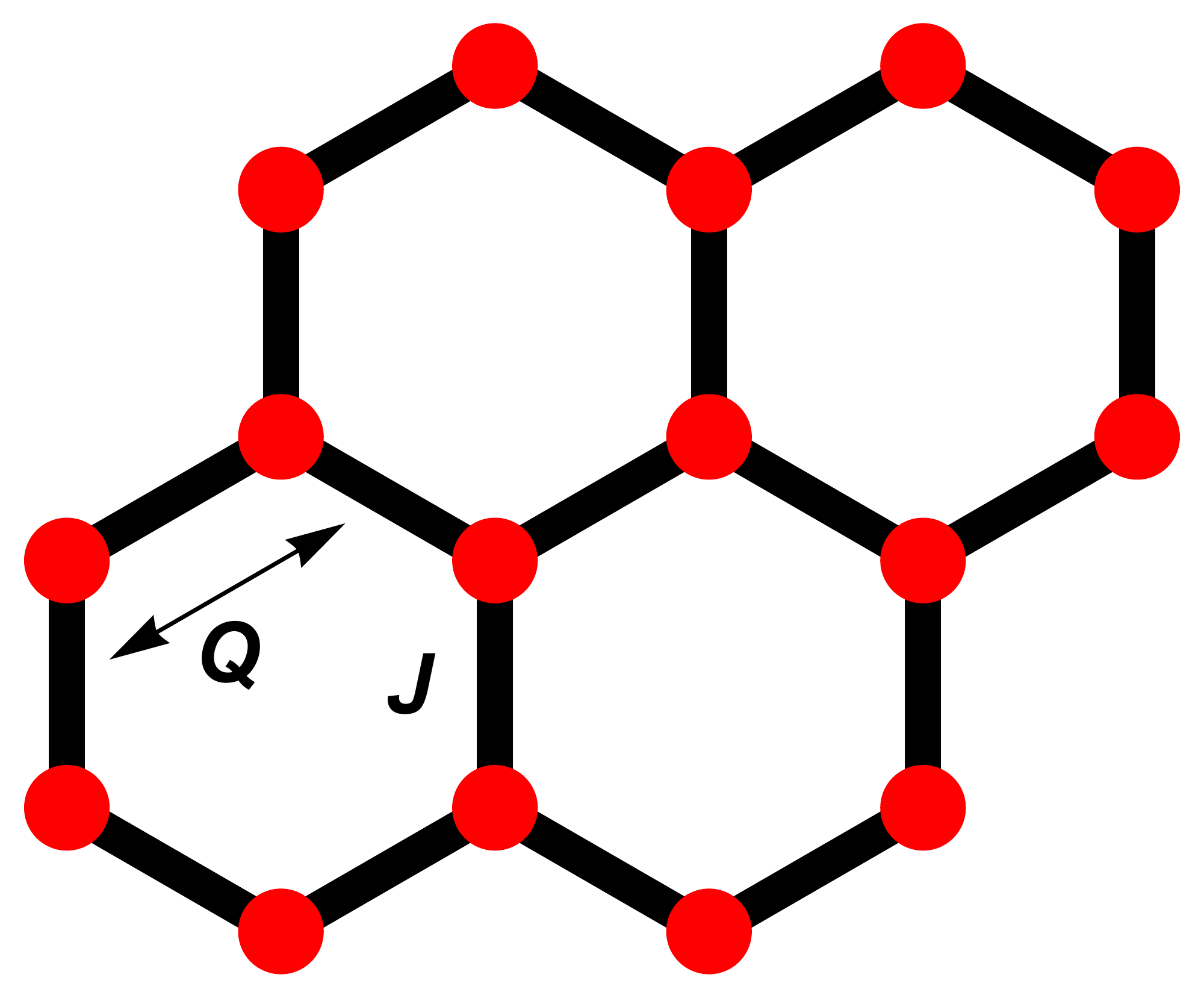}
    \caption{\label{HamiltonianChart}  The schematic representation of the model on the 2D honeycomb lattice. (a) The non-interacting part $H_0$ describes Dirac fermions at low energy, with two Dirac points located at $\pm \vec{K} = (\pm \frac{4\pi}{3},0)$. (b) For the interacting part $H_I$, $J$ is the strength of bond-bond interactions on the same bond and $Q$ represents the strength of bond-bond interactions between two different bonds of the same plaquette. }
\end{figure}

\textbf{Models}:
To investigate the nature of the transition between DSM and Kekule VBS phase with $N=1$, we introduce a generalized sign-problem-free model of $SU(N)$ fermions on the honeycomb lattice, which features a quantum phase transition between DSM and the Kekulu-VBS phases down to $N=1$. The generalized SU(N) model is given by
\bea
	H&=&H_0+H_I, 	\label{H}\\
    H_0&=&-t\sum_{\braket{ij}}\Big(c_{i}^{\dagger}c_{j}+H.c.\Big), \label{H-0}\\
	H_I&=&\!-\!\frac{J}{2N}\sum_{\braket{ij}}\Big(c_{i}^{\dagger}c_{j}+H.c.\Big)^2 \!-\!\frac{Q}{N}\sum_{\braket{ij}\braket{kl}\in P}\Delta_{ij}\Delta_{kl},~~~
	\label{H-I}
\eea
where $c^\dag_ic_j=\sum_{\sigma=1}^N c^\dag_{i\sigma}c_{j\sigma}$, $c^\dag_{i\sigma}$ creates a fermion on site $i$ with spin or flavor index $\sigma=1,\cdots$,$N$, and $\Delta_{ij}\equiv c_{i}^{\dagger}c_{j}+H.c.$ labels the hopping operator on nearest-neighbor (NN) bond $\braket{ij}$. Here $t$ is the hopping amplitude on NN bonds, $J$ is the strength of NN bond interaction, and $Q$ represents the strength of bond interactions between two next-nearest-neighbor(NNN) bonds $\braket{ij}$ and $\braket{kl}$ within the same plaquette $P$, as shown in \Fig{HamiltonianChart}. In the following, we set $t=1$ as the unit of energy. The low-energy physics of non-interacting Hamiltonian $H_0$ of the SU$(N)$ fermions at half filling in \Eq{H-0} can be described by $N$ flavors of massless four-component Dirac fermions.

When $Q=0$, the Hamiltonian in \Eq{H} is reduced to the model introduced in Ref. \onlinecite{FIQCP1}, which was shown to undergo a continuous quantum phase transition from DSM to Kekule-VBS phases with increasing value of $J$ for $N\geq 2$. However, the DSM-VBS phase transition is absent in the model with $Q=0$ and $N=1$ because the interaction $-\frac{J}{2}(c^\dag_ic_j+H.c.)^2=J(n_i-\frac12)(n_j-\frac12)$ is effectively density-density repulsion between NN sites that favors charge-density wave (CDW) order instead of Kekule-VBS at half-filling. For the $N=1$ case, a nonzero $Q$ is needed to realize a transition between the DSM and Kekule-VBS.

\begin{figure}[t]
    \includegraphics[width=.9\linewidth]{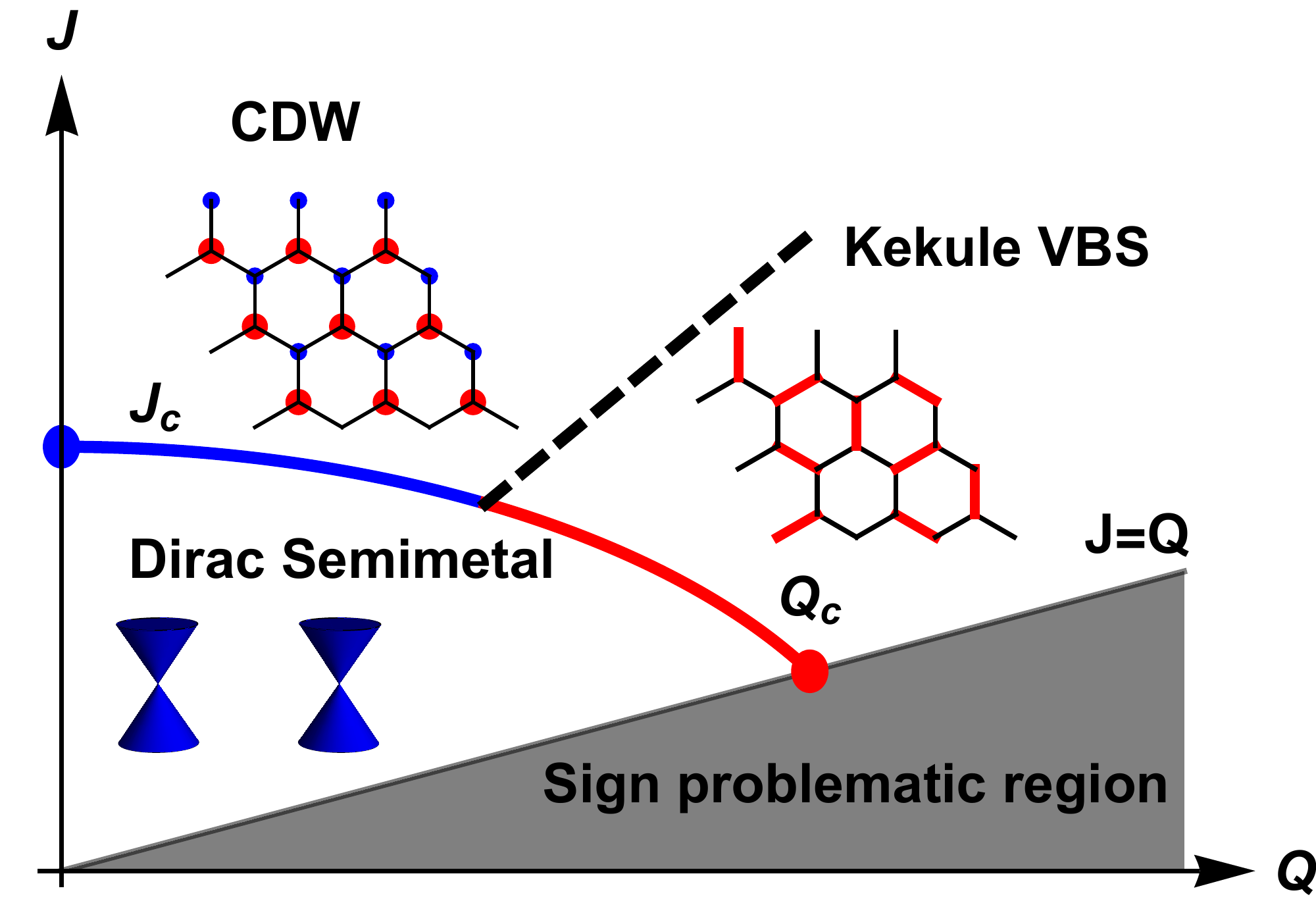}
    \caption{\label{PhaseDiagram} The schematic quantum phase diagram of the generalized model for the $N=1$ case. The shaded region is sign-problematic while for $J \geq Q>0$ the model is sign-problem-free. For $Q=0$, the quantum phase transition between DSM and CDW phase occurs at $J=J_c\approx 1.36$, consistent with results obtained in Ref. \cite{MQMC1}. Along the line of $J=Q$, our MQMC simulations show the continuous phase transition between DSM and Kekule-VBS happens at $Q=Q_c=0.16$. The phase transition between CDW and Kekule-VBS ordered phases is presumably first-order due to the incompatible symmetries of the two phases.}
\end{figure}

\textbf{MQMC simulations}: By employing the Majorana representation introduced in Ref. \onlinecite{MQMC1}, it was shown that the model at half filling with $N=1$ and $Q=0$ is sign-problem-free. For a finite $Q$, there is still a large sign-problem-free region in the $(J,Q)$ parameter space for the case of $N=1$, which allows us to perform unbiased QMC simulations to investigate the nature of quantum phase transition in systems with large lattice sizes. After rewriting $H_I$ in the following way:
\bea
&&H_I=-\frac{J-Q}{2}\sum_{\braket{ij}}(c^\dag_ic_j+H.c.)^2\nn\\
&&\!-\!\frac{Q}{2}\sum_{P} \Big[(\Delta_{i_1i_2}\!+\!\Delta_{i_3i_4}\!+\!\Delta_{i_5i_6})^2 \!+\!(\Delta_{i_2i_3}\!+\!\Delta_{i_4i_5}\!+\!\Delta_{i_6i_1})^2\Big],\nn\\
\label{H-trans}
\eea
where $i_1,\cdots,i_6$ represent six sites of hexagon plaquette $P$, it is clear that the model is sign-problem-free in the parameter region satisfying $J\geq Q\geq 0$. Hereafter, we focus on the case of $N=1$ and perform MQMC simulations in this sign-problem-free parameter region. The schematic phase diagram of the model as a function of $J$ and $Q$ is shown in \Fig{PhaseDiagram}. When the ratio of $Q/J$ is large enough, a quantum phase transition between DSM and Kekule-VBS phases can be realized with increasing $Q$. In the following, we fix $J=Q$, which is at the boundary of sign-problem-free region, and tune the value of $Q$ to explore the nature of quantum phase transition between DSM and Kekule-VBS phases.

\begin{figure}[t]
	\subfigure[]{\includegraphics[width=0.5\linewidth]{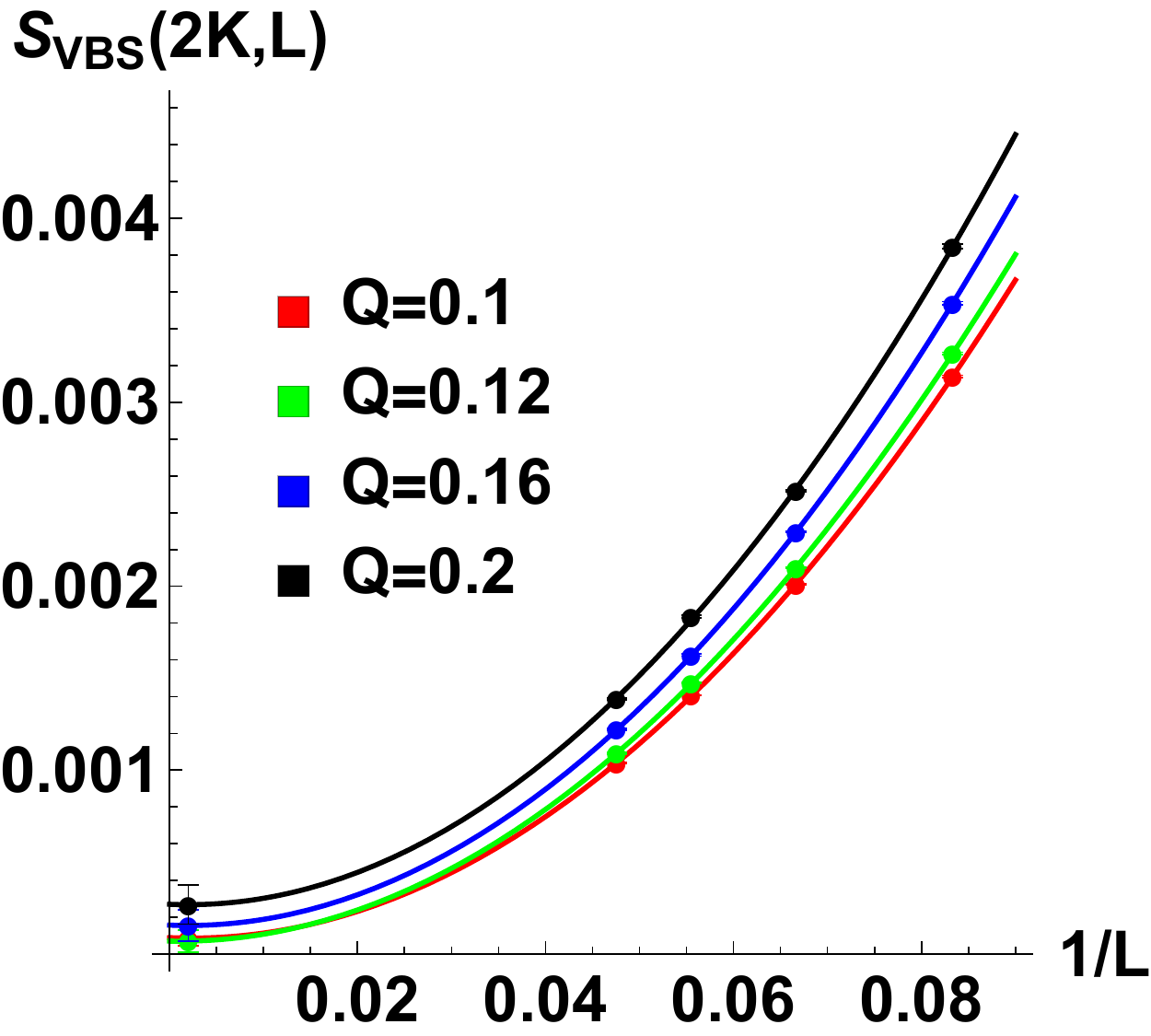}}~~
	\subfigure[]{\includegraphics[width=0.5\linewidth]{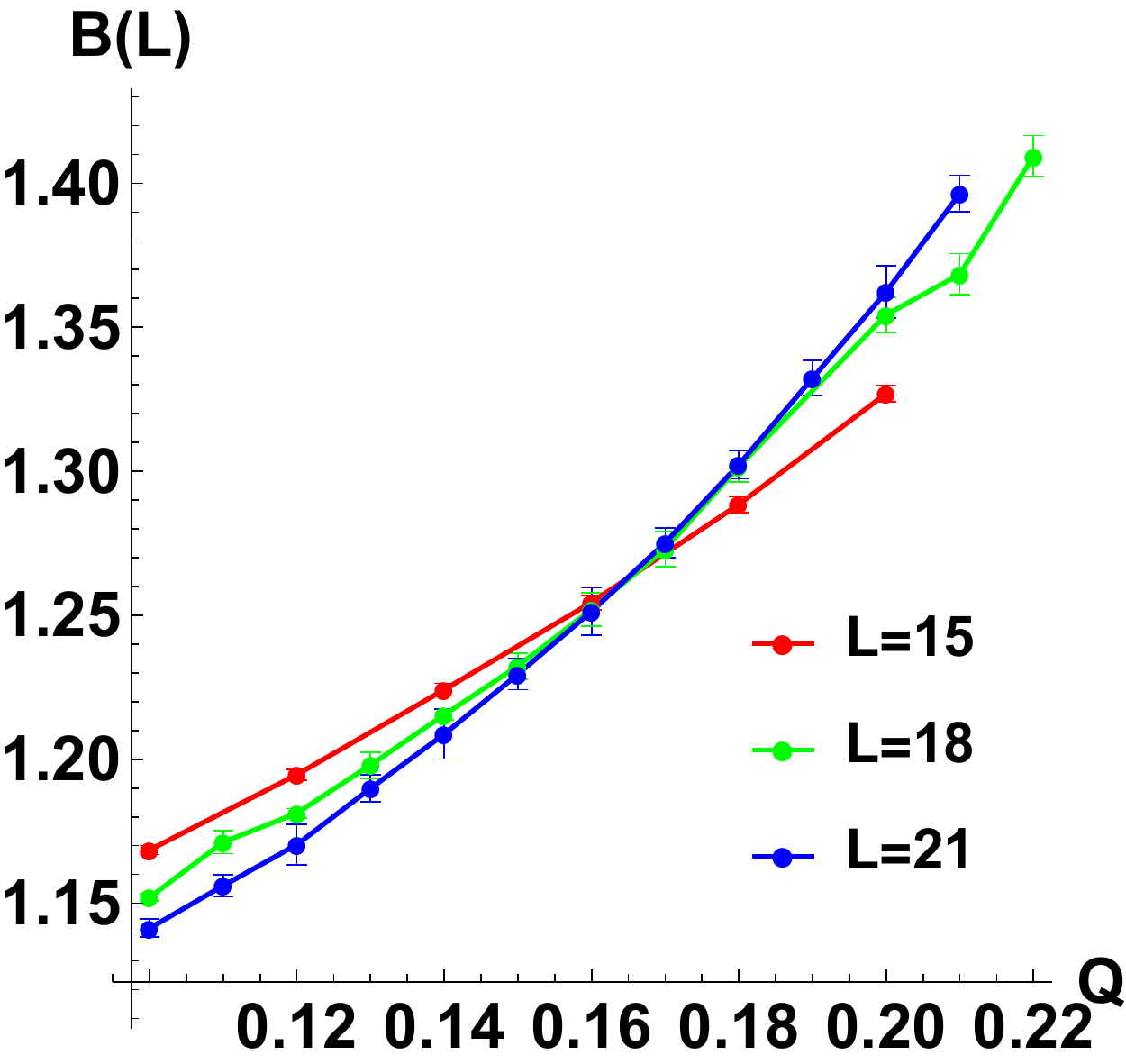}}\\
	\subfigure[]{\includegraphics[width=0.6\linewidth]{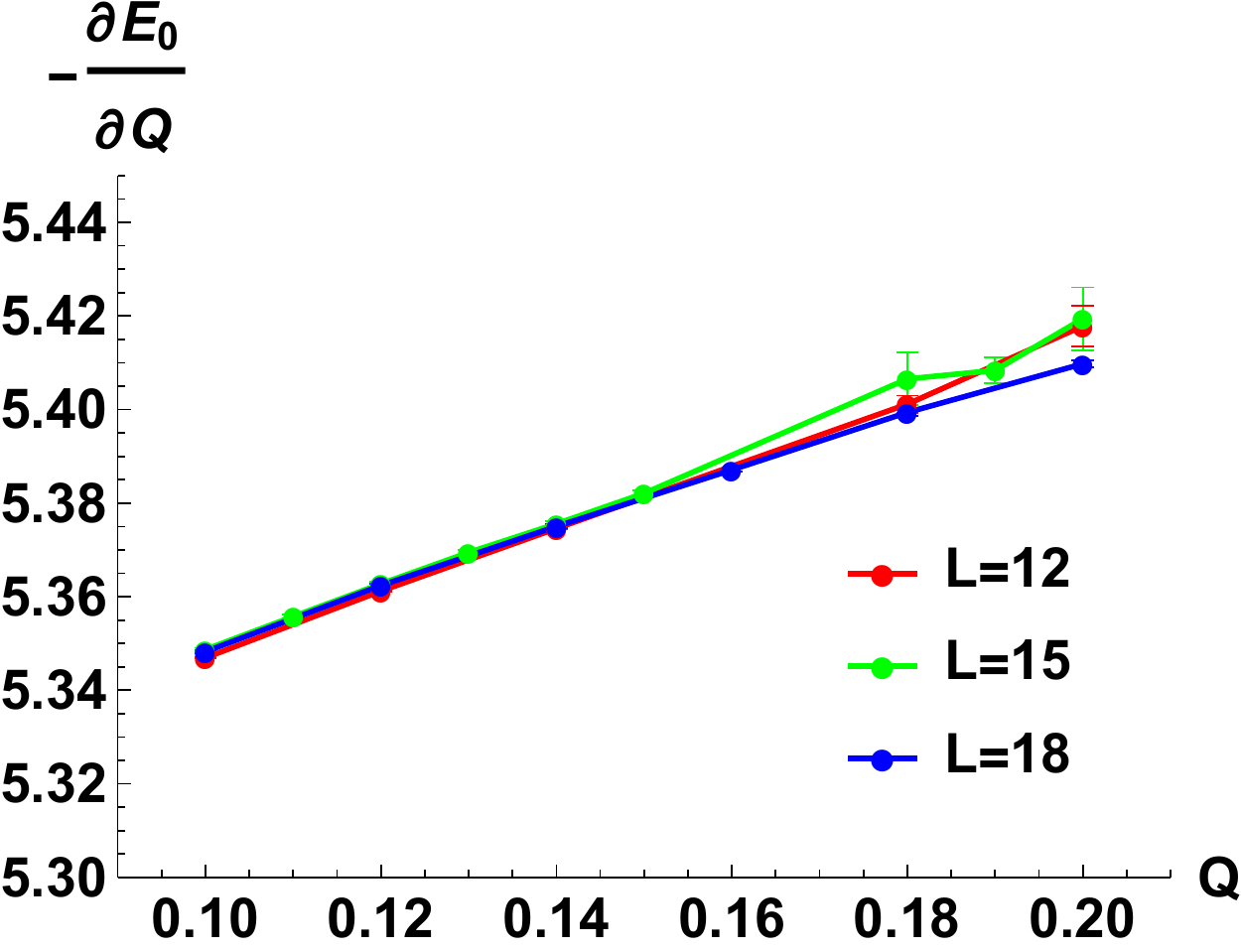}\label{FirstOrderDeri}}
    \caption{\label{QMC} MQMC simulations of the $N=1$ model \Eq{H-I} with $J=Q$. (a) Extrapolation of Kekule-VBS structure factor $S_{\textrm{VBS}}(\vec{K},L)$ with $Q$=0.10, 0.12, 0.16 and 0.20 for $N=1$ as a function of $1/L$. The Kekule-VBS order parameters are extrapolated to finite values as $L\rightarrow \infty$ when $Q$ is larger than 0.16. (b) Binder ratio $B(L)$ with different $Q$ and $L$. The phase transition occurs at $Q$=0.16. (c) The first-order derivative of the ground-state energy density with respect to $Q$. The system size in our simulation is $L=12,15,18$.}
\end{figure}

As we are interested in quantum (namely zero-temperature) phase transitions, we use projector QMC ~\cite{White1989,Sorella1990} to explore the ground-state properties of the model in Eq. \ref{HamiltonianChart}. To identify the Kekule-VBS ordering, we calculate the structure factor of VBS order parameters by MQMC: $S_{\textrm{VBS}}(\vec{k},L)=\frac{1}{L^4} \Sigma_{i,j} e^{-i\vec{k}(\vec{r_i}-\vec{r_j})}\braket{\Delta_{i,i+\delta} \Delta_{j,j+\delta}}$, where the system has $2\times L \times L$ sites with periodic boundary condition and the summation of $\delta$ over three directions of NN bonds is implicitly assumed. The Kekule-VBS order parameter $\Delta_{\textrm{VBS}}$ can be obtained through $\Delta_{\textrm{VBS}}^2=\lim_{L\rightarrow\infty}S_{\textrm{VBS}}(\vec{K},L) $ where $\vec{K}$ is the VBS ordering momentum $\pm \vec{K} = (\pm \frac{4\pi}{3},0)$ . It should be a finite value when the system lies in the Kekule-VBS phase. As shown in \Fig{QMC}(a), when the interaction is strong, namely $Q>0.16$, the VBS structure factor is extrapolated to a finite value as $L\rightarrow \infty$, indicating that the ground state possesses Kekule-VBS long-range order.

To determine accurately the critical value $Q_c$, we compute the RG-invariant Binder ratio of Kekule-VBS order $B(L)=\frac{S_{\textrm{VBS}}(\vec{K},L) }{S_{\textrm{VBS}}(\vec{K+\delta K},L) }$ for $L=9,12,15,18,21$, where $\norm{\vec{\delta K}}=\frac{2\pi}{L}$ labels minimal momentum shift from $\vec{K}$. At the putative critical point, the RG-invariant ratios of different $L$ should cross at the same point for sufficiently large $L$. The QMC results of RG-invariant ratio are shown in Fig ~\ref{QMC}(b), which clearly show that the critical point of DSM-VBS transition is about $Q_c=0.16$. The case with $Q=0$ and finite $J$ is the same as the model studied in Ref. ~\onlinecite{MQMC1,Yao2015-CDW,LeiWang2014-CDW,LeiWang2016-CDW,Wessel2016,Chandrasekharan2017}, which features a transition from DSM to CDW phases. The quantum phase transition between CDW and VBS phases, as schematic shown in \Fig{PhaseDiagram}, should be first-order since the broken symmetries of these two order parameters are incompatible with each other such that a second-order transition between these two phases is forbidden by Landau criterion (here it is not expected to feature a DQCP).

To investigate whether the transition between DSM and Kekule-VBS phase is continuous or discontinuous, we first compute the first-order derivative of the ground-state energy with respect to $Q$ in the vicinity of the transition. If a sharp kink in the derivative appears at the transition, it would indicate a first-order transition. The first-order derivative of the ground-state energy density $E_0$ with respect to $Q$ is given by
\bea
&&\frac{\partial E_0}{\partial Q} = \nonumber\\
&&\frac{-1}{2L^2} \sum_{P} \avg{(\Delta_{i_1i_2}\!+\!\Delta_{i_3i_4}\!+\!\Delta_{i_5i_6})^2 \!+\!(\Delta_{i_2i_3}\!+\!\Delta_{i_4i_5}\!+\!\Delta_{i_6i_1})^2}. \nonumber\\
\label{Partial P}
\eea
From MQMC simulations we obtain the results of first-order derivative of the ground-state energy for system size $L=12,15,18$, as shown in \Fig{FirstOrderDeri}. The tendency of discontinuity around the transition $Q_c=0.16$ is absent with all system sizes under study. It implies that the DSM-VBS transition is continuous.

\begin{figure}[t]
	\subfigure[]{\includegraphics[width=0.45\linewidth]{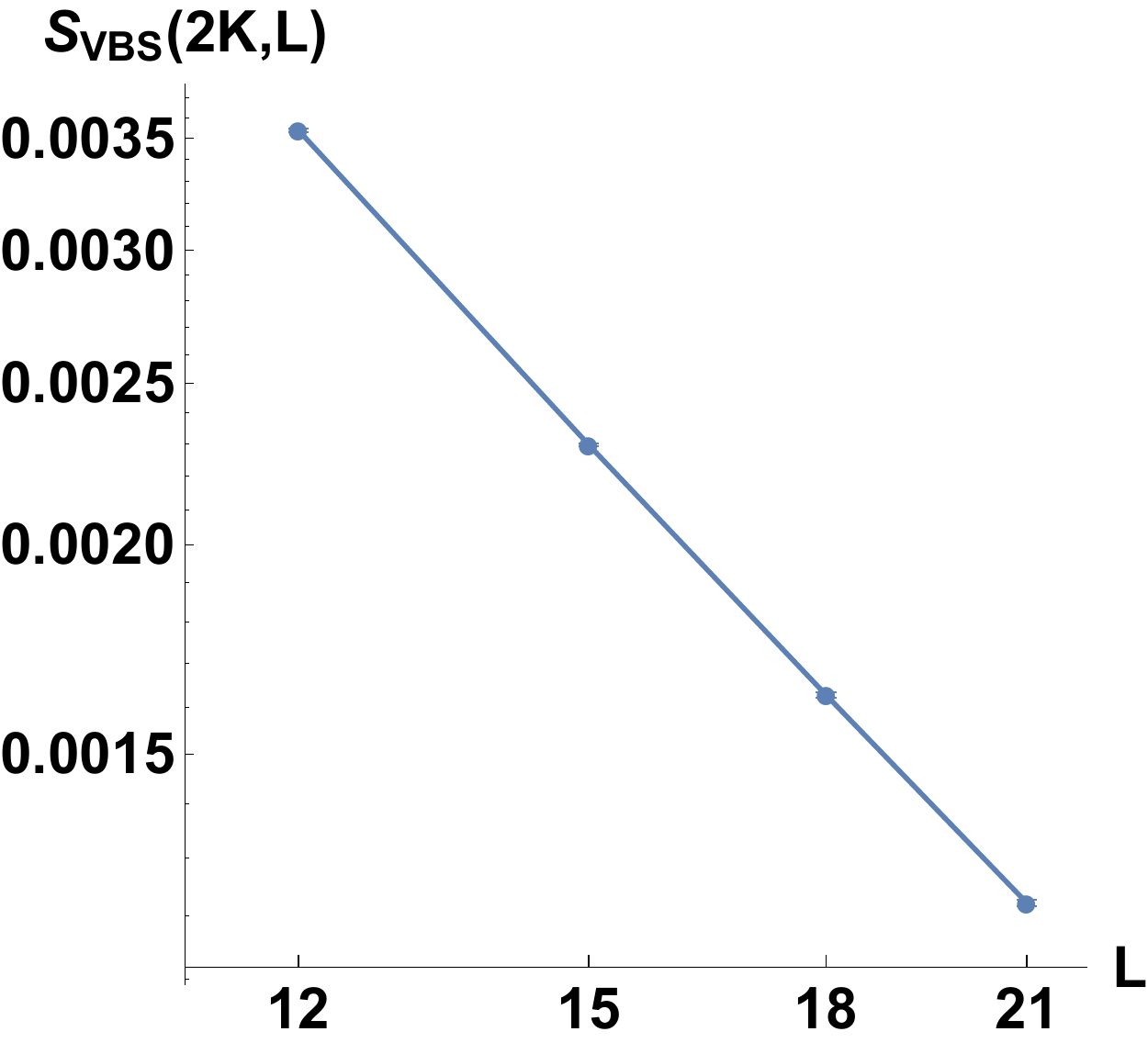}}~~
	\subfigure[]{\includegraphics[width=0.55\linewidth]{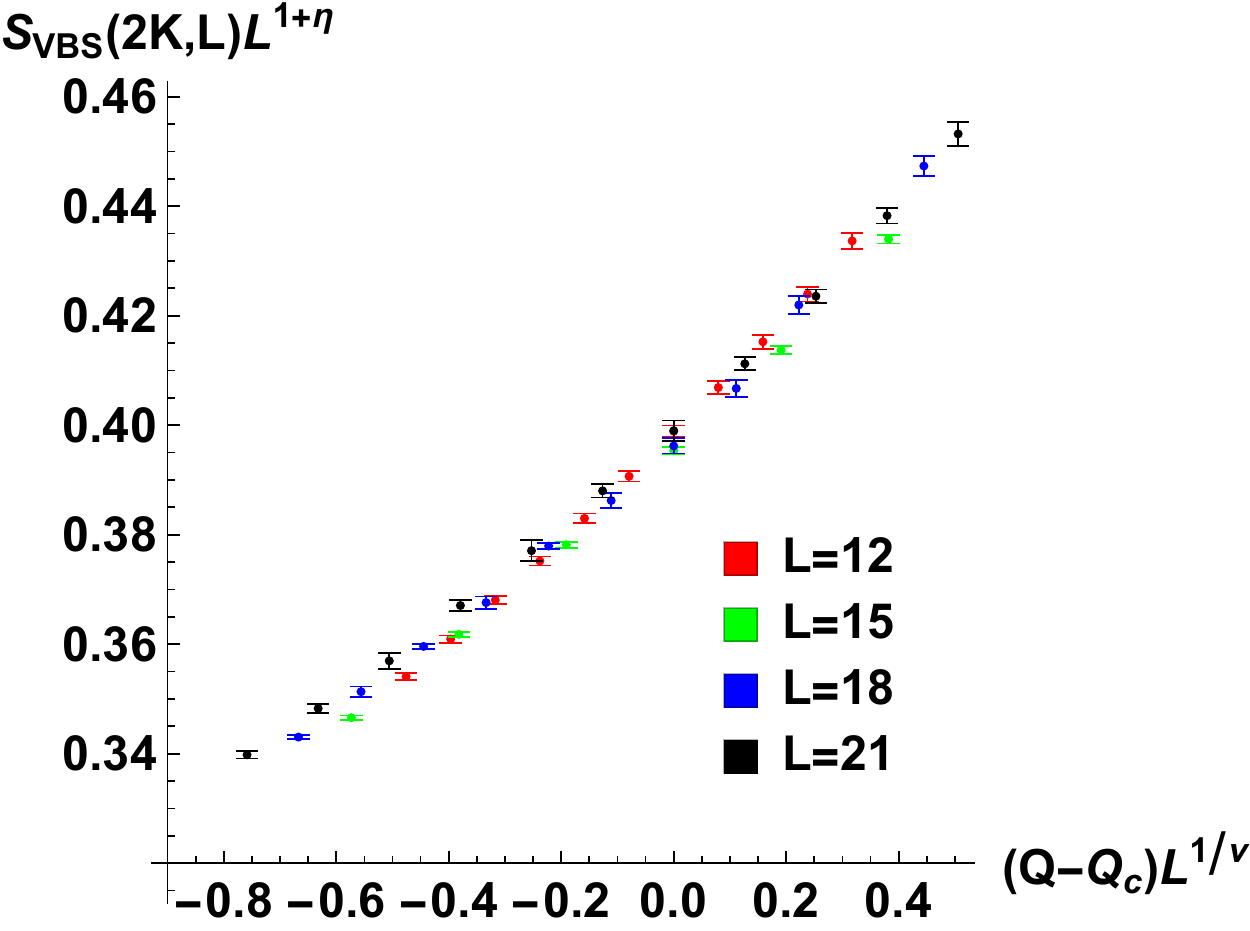}}\\
    \caption{\label{QMC2} MQMC simulation results of the $N=1$ model \Eq{H-I} with $J=Q$. (a) The anomalous dimenstion $\eta \approx 0.79(6)$ is obtained from log-log fitting of $S_{\textrm{VBS}}(\vec{K},L)$. (b) Collapsing of data points with different $Q$ and $L$ occurs when $\nu \approx 1.20$.}
\end{figure}

To better demonstrate whether the transition is continuous, we further investigate the critical behavior around the DSM-VBS transition point. More explicitly, we perform finite-size scaling analysis to extract the putative critical exponents $\nu$ and $\eta$ \cite{Sorella2016,Herbut2013-FIQCP,Toldin2015PRB}, from which other critical exponents can be obtained using hype-scaling relations. The VBS structure factors satisfy the following scaling relations with $Q$ close enough to $Q_c$ at relatively large $L$: $S_{\textrm{VBS}}(\vec{K},L)=L^{-z-\eta}\mathscr{F}(L^{\frac{1}{\nu}}(Q-Q_c))$. The dynamical exponent $z$ is assumed to be 1 due to the existence of gapless Dirac fermions at the transition point. First we obtain $\eta$ by plotting $S_{\textrm{VBS}}(\vec{K},L)$ at $Q_c$ with system size $L$ in a log-log way and fit it to a linear function with slope $-1-\eta$. With $\eta$ determined, there exists an appropriate value of $\nu$ such that data points of $(S_{\textrm{VBS}}(\vec{K},L)L^{1+\eta},L^{\frac{1}{\nu}}(Q-Q_c))$ at different $Q$ in the vicinity of $Q_c$ and different size $L$ should collapse in a single smooth curvature $\mathscr{F}$. Such finite size scaling and data collapse analysis ~\cite{Herbut2013-FIQCP,Toldin2015PRB} give rise to $\eta=0.79(6)$ and $\nu=1.20$, as shown in Fig. \ref{QMC2}(a) and (b). The collapsing of data points with different $Q$ and $L$ to a single smooth function by choosing appropriate values of $\eta$ and $\nu$ is a strong indication that the phase transition is continuous. Furthermore, the fitted result of correlation length exponent $\nu$ is much larger than $1/d$, where $d=3$ is the spacetime dimension. Such a large value of correlation length exponent $\nu$ also implies that the DSM-VBS transition should not be a first-order transition.,

We summarize the QMC results of $\eta$ and $\nu$ for $N=1$ obtained in the present work and $N=2,3,4,5,6$ of the previous work, and compare them with the results obtained in RG analysis ~\cite{FIQCP1} \cite{Herbut2017-FIQCP}, as shown in Table ~\ref{Exponents}. Our numerical results indicate that the quantum phase transition between DSM and Kekule-VBS phases for $N=1$ is continuous, namely realizing a FIQCP, which is consistent with the result of RG analysis with large-$N$ expansion. As discussed in previous works, the values of $\eta$ obtained from QMC and RG are in good agreement with each other for $N\geq 2$, especially for larger $N$. The agreement in $\nu$ is not as good as $\eta$, but still exhibits the same trend of better agreement at larger $N$. The values of $\eta$ and $\nu$ obtained from QMC for $N=1$ are slightly different from the results of RG's calculation, which might originate from the fact that RG's calculation is controlled by the parameter $1/N$ such that obtaining accurate critical exponents for $N=1$ is beyond RG's scheme. Our numerical study provides an unbiased numerical result of critical exponents for the $N=1$ Gross-Neveu chiral-XY universality class in 2+1D, which could serve as a benchmark for the higher-order RG calculation or other theoretical analysis in future studies.

\begin{table}[ttt]
    \caption{Critical Exponents $\eta$ and $\nu$ at FIQCP obtained by QMC and one-loop RG, respectively, for $N=$1,2,3,4,5,6}
	\begin{tabular}{|c|c|c|c|c|} \hline
		~~~$N$~~~ & ~~$\eta$ (RG)~~ & ~~$\eta$ (QMC)~~ & ~~$\nu$ (RG)~~ & ~~$\nu$ (QMC)~~ \\ \hline
			1 & 0.5 & 0.79(6) & 1.15 & 1.20 \\ \hline
			2 & 0.67 & 0.71(4) & 1.25 & 1.04 \\ \hline
			3 & 0.75 & 0.77(4) & 1.26 & 1.05 \\ \hline
			4 & 0.80 & 0.80(4) & 1.25 & 1.12 \\ \hline
			5 & 0.83 & 0.85(4) & 1.23 & 1.08 \\ \hline
			6 & 0.86 & 0.89(4) & 1.22 & 1.07 \\ \hline
	\end{tabular}
	\label{Exponents}
\end{table}

\textbf{Conclusion and discussion}:
In conclusion, we proposed a generalized SU($N$) fermionic model on the honeycomb lattice, which hosts the quantum phase transition between DSM and Kekule-VBS phases for the case of $N=1$ (namely the spinless fermions on the honeycomb lattice). By employing state-of-the-art MQMC simulation, we obtain convincing numerical evidences that the scenario of FIQCP, which drives the putative first-order transition to a continuous one, is realized in Dirac semimetals with flavor $N=1$. Combining with results obtained in previous works, our results indicate that FIQCP can occur in SU($N$) Dirac semimetals for all positive integers $N\geq1$ and the lower bound $N_c$ of the flavor of four-component Dirac fermions for realizing FIQCP is constrained to the range $\frac{1}{2}<N_c\leq1$. Moreover, to the best of our knowledge, our MQMC simulation is the first sign-problem-free QMC study of the critical exponents in 2+1D $N=1$ chiral-XY universality class, which provides a benchmark for the analytical calculation and numerical simulations in the future. It is also desired to derive quasi-rigorous bounds of critical exponents from conformal bootstrap calculations \cite{Ginsparg1988,Poland2019Bootstrap,Bobev2015PRL}, which is deferred to the future, and see whether the QMC results of critical exponents $\eta$ and $\nu$ are consistent with the conformal bounds.

\textit{Acknowledgments}: This work is supported in part by the NSFC under Grant No. 11825404, the MOSTC under Grant Nos. 2016YFA0301001 and 2018YFA0305604, and the Strategic Priority Research Program of Chinese Academy
of Sciences under Grant No. XDB28000000 (BHL and HY). ZXL acknowledges support from the
Gordon and Betty Moore Foundation's EPIC initiative, Grant GBMF4545. HY is also supported in part by
Beijing Municipal Science and Technology Commission (No. Z181100004218001), Beijing Natural Science Foundation
(No. Z180010), and the Gordon and Betty Moore Foundation¡¯s EPiQS Initiative through Grant GBMF4302.

\appendix
\renewcommand\thefigure{S\arabic{figure}}
\renewcommand\theequation{S\arabic{equation}}
\renewcommand\thetable{S\arabic{table}}
\setcounter{figure}{0}
\setcounter{equation}{0}
\setcounter{table}{0}

\section{\label{Assaad FSS}Finite size scaling analysis of critical properties}
In the present work, we investigate the nature of quantum phase transitions from Dirac semimetal to the Kekule-VBS phase by evaluating their structure factors in the QMC simulations, which are defined as the Fourier transform of correlation function:
\begin{equation}
S_O(\vec{k},L)=\frac{1}{L^4} \Sigma_{i,j} e^{-i\vec{k}(\vec{r_i}-\vec{r_j})}\braket{\hat{O_i}\hat{O_j}},
\label{Structure Factor}
\end{equation}
where $\hat{O}$ represents the VBS order parameter, $i,j$ are site indices, $L$ denotes the system size, and $\vec{k}$ is the crystalline momentum. For VBS order, the observable is $\hat{O_i}=c_{i\sigma}^{\dagger}c_{i+\delta \sigma}+h.c.$, where $\delta$ labels the direction
of NN bond, and the peaked momentum is $\vec{K}=(\pm \frac{4\pi}{3}, 0)$.

The RG invariant ratio, which is the ratio of structure factor defined in maintext, is a powerful tool to determine the phase transition point. In the long-range ordered phase, the RG invariant ratio $B(L)\rightarrow \infty $ for $L\rightarrow \infty$, whereas $B(L)\rightarrow 1 $ for $L\rightarrow \infty$ in the disordered phase. When system is large enough, the RG-invariant ratio is independent of system size at putative QCP as the system is scale-divergent due to the divergence of order parameter correlation length. Consequently, the phase transition point can be identified through the crossing point of RG-invariant ratio curves for different system sizes.

The critical exponents can also be extracted by structure factor and RG-invariant ratio according to their universal scaling behaviours around QCP. The universal scaling functions describing structure factor at peaked momentum and RG-invariant ratio around QCP are:
\bea
	&&S(\vec{K},L)=L^{-(d+z-2+\eta)}\mathscr{F}_1(L^{\frac{1}{\nu}}(Q-Q_c))	\nonumber\\
		&&B(L)=\mathscr{F}_2(L^{\frac{1}{\nu}}(Q-Q_c)),
	\label{Critical Exponent}
\eea
where $d$ represents the spatial dimension and $\vec{K}$ is the peak momentum of VBS structure factor. The critical exponent $\eta$ is anomalous dimension and $\nu$ is the correlation function exponent. $z$ is dynamical critical exponent, which has the value $z=1$ due to Dirac physics. $\mathscr{F}_1$ and $\mathscr{F}_2$ are unknown ansatz scaling functions. Based on the above scaling function, we can extract the critical exponent $\eta$:
\bea
\eta(L)=
\frac{1}{\frac{\log(L)}{\log(L+3)}} \log\frac{S(\vec{K},L+3)}{S(\vec{K},L)} \vert_{Q=Q_c(L)} -(d+z-2),\nn\\
\label{Eta L}
\eea
where $\eta(L)$ is the anomalous dimension $\eta$ extracted at the crossing point $P_C(L)$ of $B(L)$ and $B(L+3)$.

\begin{figure}[t]
    \includegraphics[width=0.8\linewidth]{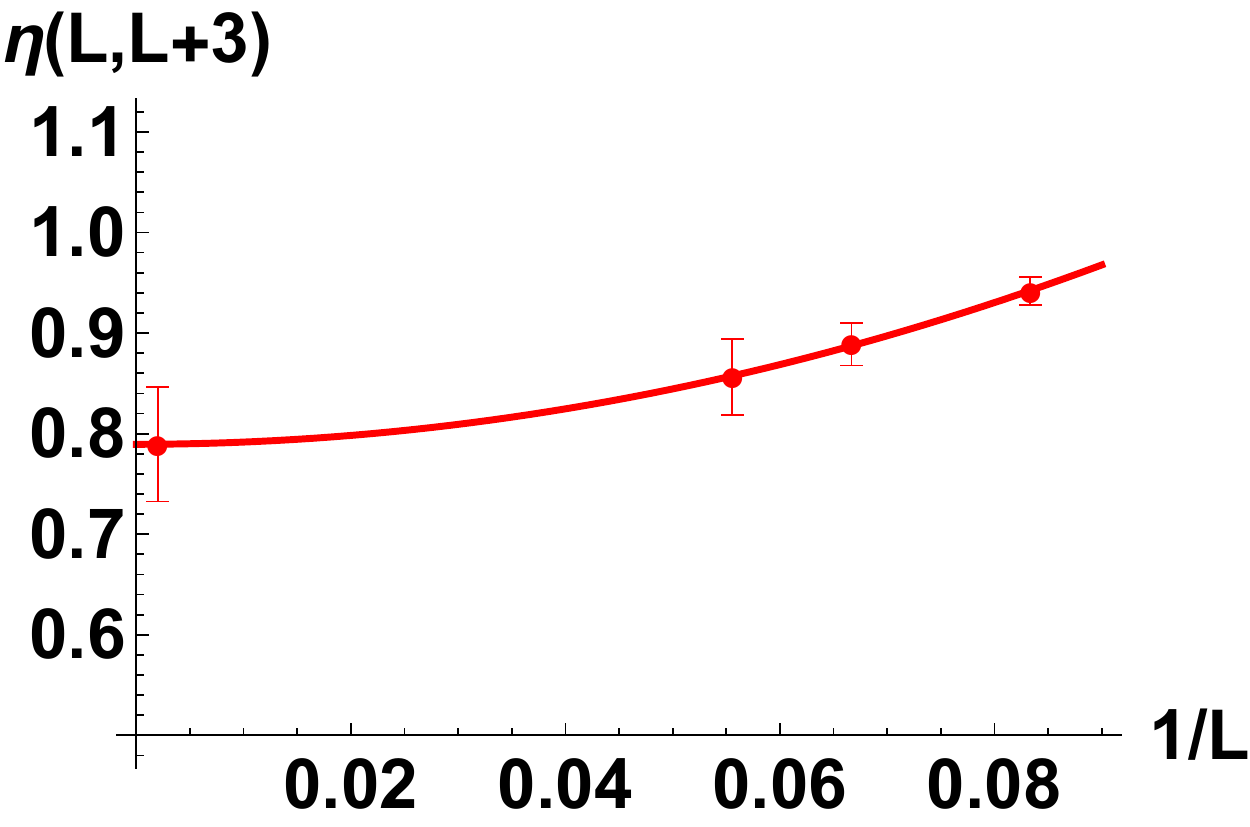}
    \caption{\label{AssaadEta}The extrapolation of the anomalous dimension $\eta$ to the thermodynamic limit for Kekule-VBS phase transitions. The system size in our simulation is $L = 12,15,18$. The extrapolated $\eta(L\rightarrow \infty)$ at thermaldynamical limit is approximately 0.79(6). }
\end{figure}

\section{Renormalization Group analysis of DSM-VBS transition for $N=1$ }

The low-energy field theory describing the quantum phase transitions between DSM and Kekule-VBS phases has been constructed in previous works ~\cite{FIQCP1}. Near the quantum phase transition, the system can be described by $SU(N)$ Dirac fermions $\psi$, fluctuating $Z_3$-order parameters $\phi$, and their couplings: $S=S_{\psi}+S_{\phi}+S_{\psi \phi}$. The action $S_\psi$ for $SU(N)$ Dirac fermions on honeycomb lattice is given by:
\bea
S_{\psi}=\int d^3 x \psi^{\dagger}[\partial_{\tau}-v(i \sigma^x \tau^z \partial_x+ i \sigma^y \tau^0 \partial_y)]\psi,\nn
\label{actionpsi}
\eea
where $\tau^i$ ($\sigma^i$) are Pauli matrices on valley (sublattice) spaces, $v$ denotes the Fermi velocity and $\psi=(\psi_{\vec{K}A}^{\dagger}(x), \psi_{\vec{K}B}^{\dagger}(x), \psi_{-\vec{K}A}^{\dagger}(x), \psi_{-\vec{K}B}^{\dagger}(x))$ is the four-component fermion creation operator with $\pm \vec{K} = (\pm \frac{4\pi}{3},0)$ denoting valley momenta of Dirac points and $A,B$ labelling sublattice. The flavor index $\sigma=1,2,3, ..., N$ is implicit in the action. For spin-$\frac{1}{2}$ fermions on the honeycomb lattice (e.g. graphene), $N=2$. For spinless fermions on the honeycomb lattice, $N=1$, which is the focus of the present work.

\begin{figure}[t]
   \includegraphics[width=0.83\linewidth]{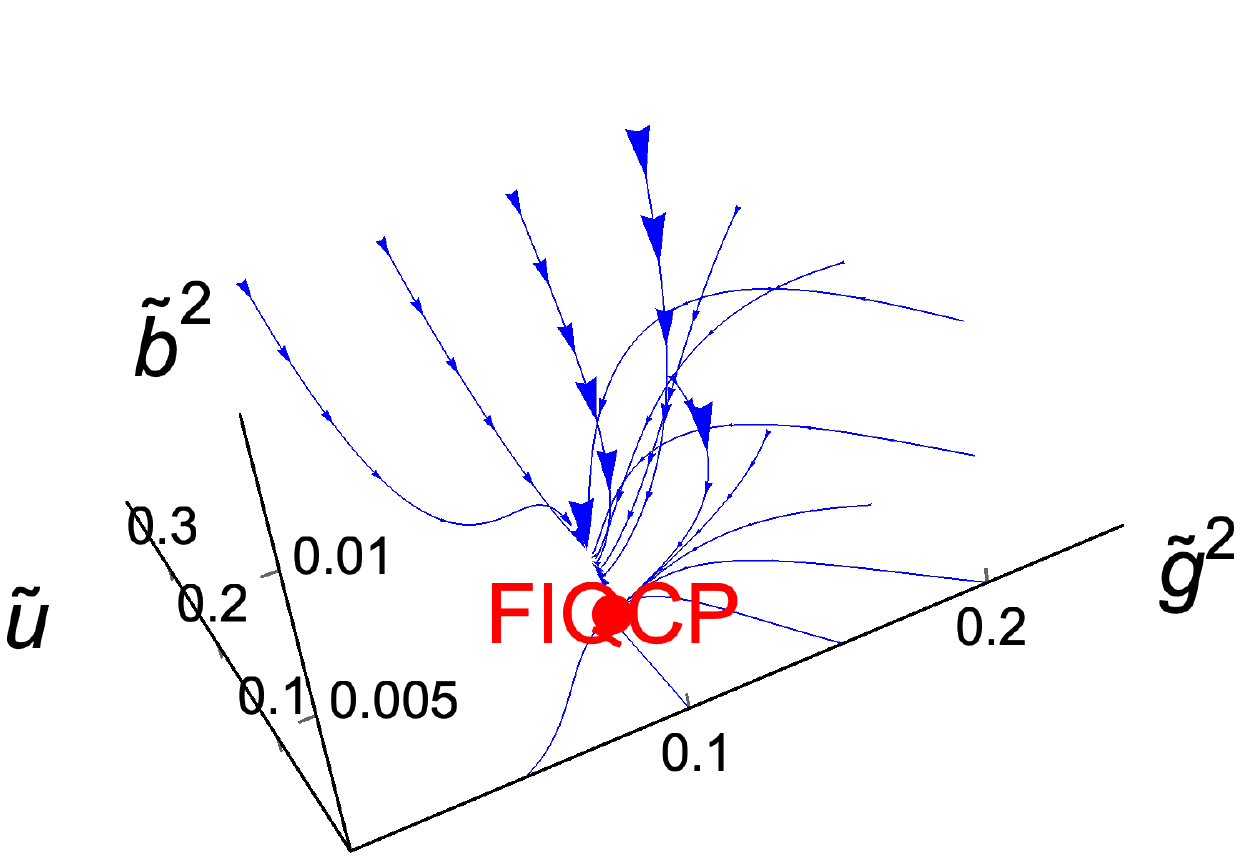}
   \caption{\label{RGflow}RG flow of coupling constants in the critical hypersurface $r=r_c$ at one-loop level for $N=1$, according to RG equations obtained in Ref. \onlinecite{FIQCP1}. The only stable fixed point denoted by the red point is the Gross-Neveu-Yukawa (GNY) fixed point representing a FIQCP.}
\end{figure}

The Kekule-VBS ordering breaks the lattice translation symmetry with wave vectors $\pm 2\vec{K}$ and $C_3$ rotational symmetry. The action effectively describing the fluctuations of order-parameter bosons is given by
\bea
S_{\phi}\!=\!\int d^3 x\big[\norm{\partial_{\tau}}^2+c^2 \norm{\nabla \phi}^2+r\norm{\phi}^2+b(\phi ^3 + \phi ^{*3})+u \norm{\phi}^4\big],\nn
\label{actionphi}
\eea
where $\phi(x)=\phi_{2\vec{K}}(x)$ is the complex-valued order-parameter and ${r,c,b,u}$ are real constants. Note that the cubic term with coefficient $b$ is allowed by symmetry in the action above. According to the Landau cubic criterion, the cubic term should drive the phase transition first-order. However, as shown in Ref. \onlinecite{FIQCP1} and Ref. \onlinecite{FIQCP3}, the coupling between Dirac fermions and fluctuating order-parameters
\bea
S_{\psi \phi}=g\int d^3 x (\phi \psi^{\dagger}\sigma^x \tau^+ \psi +h.c.),\nn
\label{actioncoupling}
\eea
where $\tau^{\pm}=(\tau^x \pm i \tau^y)/2$ and $g$ labels Yukawa coupling strength, can qualitatively affect the nature of the $Z_3$ transition and may render this putative first-order transition into a continuous one.

The large-$N$ RG analysis of the transition between DSM and Kekule-VBS described by the action $S$ above was performed in Ref. \onlinecite{FIQCP1}. By solving the RG flow equations describing the flow of coupling constants upon integrating out fast modes, it was shown that for $N>1/2$ there is only one stable fixed point: $(\tilde{g}^2, \tilde{b}^2, \tilde{u})\approx (\frac{2}{\pi N},0,\frac{2}{\pi N})$ on the critical surface ($r=r_c$), where the dimensionless constants $(\tilde{g}^2,\tilde{b}^2, \tilde{u})$ were obtained from $(g^2,b^2,u)$ upon rescaling. In other words, for $N>1/2$ the cubic term $b$ is irrelevant and the putative first-order DSM-VBS transition is induced to a continuous one. The RG flow for the case of $N=1$ is shown in \Fig{RGflow}. When $N>1/2$, by solving the linearized RG equations in the vicinity of the Gross-Neveu-Yukawa (GNY) fixed point, the critical exponents are given by $\eta=\frac{N}{N+1}$ and $\nu=2-\frac{1+4N+\sqrt{1+38N+N^2}}{5(1+N)}$.
For $N<1/2$, the cubic term $b$ is shown from the large-$N$ analysis to be relevant and features a run-away flow. Consequently, the large-$N$ RG analysis at the one-loop level predicted a critical $N_c=\frac{1}{2}$. Here we would like to mention that the scaling dimension of the cubic term is exactly 2 at the $N=1/2$ GNY fixed point with emergent supersymmetry, which is less than spacetime dimension 2+1, implying that the cubic term is relevant and the DSM-Kekule VBS transition of $N=1/2$ should be a first-order transition, as shown in Ref. \onlinecite{FIQCP1}. Because it is a first-order transition at $N=1/2$, we can conclude that the exact value of $N_c$ should obey $N_c>\frac{1}{2}$.

\bibliography{FIQCPN1_Oct28Bohai}
\end{document}